\documentclass{wqcd03}                 

\usepackage{txfonts}                   

\newcommand{\beq}{\begin{equation}}
\newcommand{\eeq}{\end{equation}}
\def\ord{{\cal O} }
\newcommand\sss{\scriptscriptstyle}
\newcommand\as{\alpha_{\sss S}}

\newcommand\mh{M_{\sss {\rm H}}}
\newcommand\mt{M_{\rm  top}}
\def\muf{\mu_{\sss F}}
\def\mur{\mu_{\sss R}}
\newcommand\sah{s_{j_1\sss {\rm H}}}
\newcommand\sbh{s_{j_2\sss {\rm H}}}
\newcommand\ptj{p_{{\sss T} j}}

\confname{QCD@Work 2003 - International Workshop on QCD, Conversano, Italy, 
14--18 June 2003}
\title{Higgs Production at LHC}

\author{Vittorio Del Duca\addressmark{a} }

\address[a]{Istituto Nazionale di Fisica Nucleare, sez. di Torino, 
via P. Giuria 1, 10125 - Torino, Italy}

\begin{document}

\begin{abstract}
We review Higgs production at hadron colliders via vector-boson fusion and 
via gluon fusion,
fully inclusively and in association with one and two jets.
Then we address the issue of the measurement of the coupling of the Higgs
to the vector bosons at LHC.
\end{abstract}

\maketitle


\section{Introduction}

The mechanism that governs the electroweak symmetry breaking is at
present the largest unknown in the Standard Model (SM) of elementary
particle physics. The canonical mechanism, the Higgs model, is a
keystone of the SM and its supersymmetric extensions. However, it is
based on the existence of a CP-even scalar particle, the Higgs boson,
which has not been detected yet and is the most wanted particle of the
Fermilab Tevatron and the CERN Large Hadron Collider (LHC) physics
programmes.  The direct search in the $e^+e^-\to ZH$ process at the
CERN LEP2 collider has posed a lower bound of 114.1~GeV on the SM Higgs mass, 
$\mh$~\cite{Barate:2000ts,Acciarri:2000hv,Abbiendi:2000ac,Abreu:2000fw,:2001xw}.
LEP2 also posed lower bounds of 91.0~GeV (91.9~GeV)
on the CP-even (CP-odd) Higgs bosons of the minimal supersymmetric
extension of the Standard Model (MSSM)~\cite{unknown:2001xx}.
On the other hand, a SM Higgs boson with a mass of the order of 200~GeV
or less is favoured by analyses of the electroweak precision
data~\cite{Gambino:2003xc}, and the upper bound on the
lightest Higgs boson in the MSSM is at about
135~GeV~\cite{Carena:2000dp,Espinosa:2000df,Brignole:2001jy}.  Thus the
intermediate mass region ($\mh \lesssim$ 200 GeV) seems to be preferred.

In $pp$-collisions,
the dominant production mechanism over the entire Higgs mass range
accessible at LHC is via gluon fusion $g g\to H$, where the coupling of the
Higgs to the gluons is mediated by a heavy quark loop.
If the mass of the Higgs is light ($\mh \lesssim$ 150 GeV),
as one may expect from the electroweak fits to the LEP data,
the most important decay mode is the rare decay into two photons,
$H\to\gamma\gamma$~\cite{cms,atlas}, characterised by a very narrow mass peak.
In inclusive searches, the main Higgs decays
$H\to b \bar{b}\,, \tau^+\tau^-$ are 
overwhelmed by large QCD backgrounds.

The second largest production mechanism is via weak-boson fusion 
(WBF) $q q\to q q H$. In terms of significance,
{\it i.e.} ratio of signal over root of background $S/\sqrt{B}$,
Higgs production via WBF is large as (or larger than) the one
via gluon fusion~\cite{atlas2}. In addition, it allows for the study of
the Higgs couplings to gauge bosons and fermions~\cite{Zeppenfeld:2000td} 
at LHC, because of the coupling to the weak bosons in the production mode and
because of the different decay modes which can observed 
experimentally~\cite{atlas2},
$H\to\gamma\gamma$~\cite{wbfhtophoton}, $H\to WW$~\cite{wbfhtoww,Kauer:2000hi}
and $H\to\tau\tau$~\cite{wbfhtautau,Plehn:1999xi}. 

Firstly, we review in general Higgs production via WBF and via gluon fusion,
and then we consider both in association with two jets, as regards the 
Higgs-vector-boson couplings.

\section{Higgs Production via Vector Boson Fusion}

\begin{figure}[h]
\hbox to\hsize{\hss
\includegraphics[width=.5\hsize]{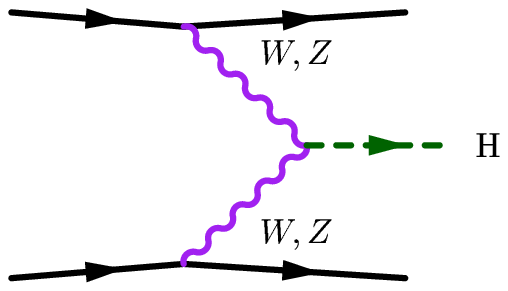} 
\includegraphics[width=.5\hsize]{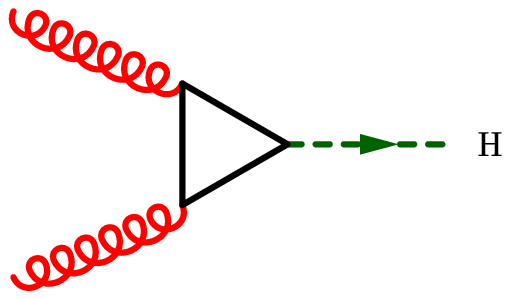}
\hss}
\caption{Higgs production via WBF and via gluon fusion.}
\label{fig:fig1}
\end{figure}

Higgs production via WBF $q q\to q q H$ occurs as the scattering between
two (anti)quarks with weak-boson ($W$ or $Z$) exchange in the $t$-channel
and with the Higgs boson radiated off the weak-boson propagator, 
Fig.~\ref{fig:fig1}. Since the distribution functions of the incoming valence
quarks peak at values of the momentum fractions 
$x\approx 0.1$ to 0.2, Higgs production via WBF tends to produce naturally two
highly energetic outgoing quarks. In addition, since the weak-boson mass 
provides a natural cutoff on the weak-boson propagator, two jets with
a transverse energy typically of the order of a fraction of the weak-boson 
mass can be easily produced. Thus a large fraction of the events of the
total cross section for Higgs production via WBF occurs with two jets with
a large rapidity interval between them, typically one at forward and the
other at backward rapidities. As we shall see, this feature can be used
in order to distinguish Higgs production via WBF from the one via gluon
fusion. 

Another feature of WBF
is that, except for the Higgs decay products, which may be hadronic,
no hadron production occurs in the rapidity interval between the jets,
because the weak boson exchanged in the $t$-channel is 
colourless~\cite{Dokshitzer:1987nc,Bjorken:1993er}:
to $\ord(\as)$, gluon radiation occurs only as bremsstrahlung off the quark
legs (since no colour is exchanged in the $t$-channel in the Born process,
no gluon exchange is possible to $\ord(\as)$, except for
a tiny contribution due to equal-flavour quark scattering with 
$t\leftrightarrow u$ channel exchange). Next-to-leading order (NLO) corrections
in $\as$ to Higgs production via WBF have been computed for the total cross 
section~\cite{WBF_NLO} and for Higgs production in association with two 
jets~\cite{Figy:2003nv}. They have been found to be typically modest, 
of the order of 5 to 10\%. Thus Higgs production via WBF is under a good
theoretical control.

\section{Higgs Production via Gluon Fusion}

\begin{figure}[h]
\hbox to\hsize{\hss
\includegraphics[width=\hsize]{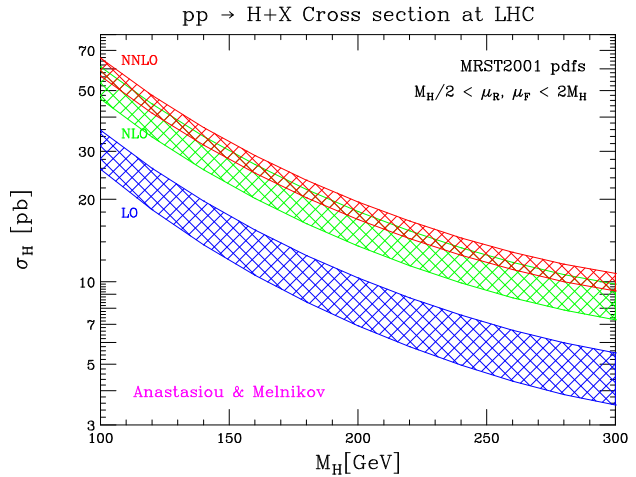}
\hss}
\caption{Higgs production via gluon fusion in $p p$ collisions at
$\sqrt{S} = 14$~TeV as a function of the Higgs mass, from 
Ref.~\cite{Anastasiou:2002yz}. The production rate has been computed 
in the large $\mt$ limit, to leading order, NLO and NNLO accuracy. 
The shaded bands 
display the renormalisation $\mur$ and factorisation $\muf$ scale variations.
The lower contours correspond to $\mur = 2\mh$ and $\muf = \mh/2$;
the upper contours to $\mur = \mh/2$ and $\muf = 2\mh$.}
\label{fig:fig2}
\end{figure}

In the gluon fusion channel $g g\to H$, the Higgs couples to the gluons 
through a heavy quark loop, Fig.~\ref{fig:fig1}. In the SM,
the leading contribution comes from the top quark, the
contributions from other quarks being at least smaller by a factor
$\ord(M_{\rm b}^2/\mt^2)$.  Since the Higgs boson is produced via a
loop, a calculation of the production rate is quite involved,
even at leading order in $\as$.  The production rate for $g g\to H$ has
been computed to NLO in $\as$, including the
heavy quark mass dependence~\cite{Graudenz:1992pv,Spira:1995rr} 
(which required an evaluation at two-loop accuracy).  The NLO QCD
corrections are large and increase the production rate by up to 80\%.
However, the coupling of the Higgs to the gluons via a top-quark loop
can be replaced by an effective coupling~\cite{Shifman:1979eb,Ellis:1976ap},
called the {\it large $\mt$ limit}, if the Higgs mass is smaller than
the threshold for the creation of a top-quark pair, $\mh \lesssim 2 \mt$.
That simplifies calculations tremendously, because it effectively
reduces the number of loops in a given diagram by one.  It has been
shown that adding the NLO QCD corrections in the large $\mt$ limit
to the leading order calculation with the top quark mass dependence 
approximates the full NLO QCD corrections within 10\,\% up to
1~TeV~\cite{Kramer:1996iq} covering the entire Higgs mass range 
at the LHC.  The reason for the good quality of this approximation is that
the QCD corrections to $g g\to H$ are dominated by soft gluon effects,
which do not resolve the top-quark loop mediating the coupling of the
Higgs boson to the gluons. The next-to-next-to-leading order (NNLO)
corrections to the production rate for $g g\to H$ have been evaluated
in the large $m_t$ 
limit~\cite{Harlander:2002wh,Anastasiou:2002yz,Ravindran:2003um}
and display an increase of about 15\%\ at $\mh = 120$~GeV
with respect to the NLO evaluation, Fig.~\ref{fig:fig2}. 
The dominant part of the NNLO corrections comes from the gluon and
collinear radiation~\cite{Catani:2001ic,Harlander:2001is}, in agreement
with what already observed at NLO. In addition, the threshold
resummation of soft gluon effects~\cite{Kramer:1996iq,Catani:2003zt}
enhances the NNLO result by about 5\%, showing that the calculation
stabilises at NNLO.

For a light Higgs ($\mh \lesssim$ 150 GeV), 
the most relevant decay mode is into two photons,
$H\to\gamma\gamma$. 
The irreducible $p p\to\gamma\gamma$ background proceeds at leading
order via the sub-process $q\bar{q}\to\gamma\gamma$, which is
independent of $\as$. The NLO corrections to $p p\to\gamma\gamma$ are 
known~\cite{Aurenche:1985yk,Bailey:1992br,Bailey:1992jz}.  They are
incorporated in the program DIPHOX, which includes all relevant photon
fragmentation effects~\cite{Binoth:1999qq}.  The NLO corrections are
very large, because the formally higher order sub-processes involving
gluons in the initial state can be as large as the leading order
contribution, since the gluon distribution in the proton decreases
monotonically with its momentum fraction $x$. Thus the
$q g\to\gamma\gamma q$, occurring first to NLO, dominates the NLO
corrections.  In fact, the quark-loop mediated sub-process 
$g g\to\gamma\gamma$~\cite{Ametller:1985di,Dicus:1987fk}, which is
$\ord(\as^2)$ and thus formally belongs to the NNLO corrections,
contributes about 50\% to the $\gamma\gamma$
background~\cite{deFlorian:1999tp}.
However, since $g g\to\gamma\gamma$ occurs first at $\ord(\as^2)$, it
is effectively a leading order calculation to that order, and thus it
bears a large dependence on renormalisation and factorisation scales.
In order to reduce that uncertainty, the NLO corrections to
$g g\to\gamma\gamma$ (which are $\ord(\as^3)$ and thus formally belong
to the NNNLO corrections) have been evaluated~\cite{Bern:2002jx}.  They
were found to be modest, increasing the overall irreducible
$p p\to\gamma\gamma$ background by less 10\% over the relevant diphoton
mass spectrum, $m_{\gamma\gamma}\gtrsim$~100~GeV.  Other NNLO
corrections to the irreducible $p p\to\gamma\gamma$ background, like
the sub-process $g g\to\gamma\gamma q\bar{q}$ and the $\ord(\as)$
corrections to $q g\to\gamma\gamma q$, might be numerically relevant,
so a full NNLO evaluation of the irreducible $p p\to\gamma\gamma$
background is desirable.

\subsection{Higgs $+$ 1 Jet Production}

For a light Higgs,
a process that promises to have a more amenable background is Higgs
production in association with a high transverse energy ($E_{\sss T}$)
jet, $pp\to H +$ jet $\to \gamma\gamma +$ jet. In addition, this process
offers the advantage of being more flexible with respect to choosing
suitable acceptance cuts to curb the background.  The $pp\to H +$ jet
process is known to leading order exactly~\cite{Ellis:1988xu}.
The contributing subprocesses include 
quark-gluon scattering which is mediated by top-quark triangles, 
and gluon-gluon scattering which requires box diagrams.
As regards the NLO contribution, only the bremsstrahlung corrections
are known~\cite{DelDuca:2001eu,DelDuca:2001fn}. However, the full NLO 
corrections~\cite{deFlorian:1999zd,Ravindran:2002dc,Glosser:2002gm,mcfm}
have been evaluated in the large
$\mt$ limit. For Higgs + one jet production, the large $\mt$
limit is valid as long as $\mh \lesssim 2 \mt$ and the transverse
energy is smaller than the top-quark mass, 
$E_{\sss T}\lesssim \mt$~\cite{Baur:1989cm}, and is insensitive
to the jet-Higgs invariant mass becoming larger than 
$\mt$~\cite{DelDuca:2003ba}. At $\mh = 120$~GeV,
the NLO corrections to the $pp\to H +$ jet process increase the leading
order prediction by about 60\%, and thus are of the same order
as the NLO corrections to fully inclusive $pp\to H$ production
considered above.  At present, the NNLO corrections to $pp\to H +$ jet
are not known.  

The irreducible $p p\to\gamma\gamma$ jet background has been analysed
at leading order in Ref.~\cite{Abdullin:1998er}. It
proceeds via the sub-processes $q\bar{q}\to\gamma\gamma g$ and
$q g\to\gamma\gamma q$, which are $\ord(\as)$, and 
is dominated by $qg\to \gamma\gamma q$,
which benefits from the large gluon luminosity.  The quark-loop
mediated $gg\to g \gamma\gamma$ sub-process, which is $\ord(\as^3)$ and
thus formally belongs to the NNLO corrections, yields a contribution 
to $p p\to\gamma\gamma$ jet which increases the leading order
prediction by less than 20\%~\cite{deFlorian:1999tp,Balazs:1999yf}, thus
it is under good theoretical control. The
full NLO QCD corrections to the irreducible $pp\to\gamma\gamma$ jet
background have been computed~\cite{DelDuca:2003uz} using a ``smooth'' photon
isolation prescription~\cite{Frixione:1998jh} which does not require a photon
fragmentation contribution. They have been found to depend strongly on
the photon isolation parameters. In particular, choosing a small photon 
isolation cone radius $R_\gamma = 0.4$ (which is nowadays the experimental 
preferred choice) results in more than 100\,\% corrections.

\section{Higgs $+$ 2 Jet Production}

A key component of the program to measure Higgs boson couplings at the LHC is
the WBF process, $qq\to qqH$ via $t$-channel $W$ or $Z$
exchange, characterized by two forward quark jets~\cite{Zeppenfeld:2000td}.
The NLO corrections to Higgs production via WBF fusion in association with
two jets are known to be small~\cite{Figy:2003nv} and,
hence, this process promises small systematic errors. $H+2$~jet production
via gluon fusion, while part of the inclusive Higgs signal, constitutes a
background when trying to isolate the $HWW$ and $HZZ$ couplings responsible
for the WBF process. A precise description of this background is needed in
order to separate the two major sources of $H+2$~jet events: one needs to
find characteristic distributions which distinguish the WBF
process from gluon fusion.

\begin{figure}[h]
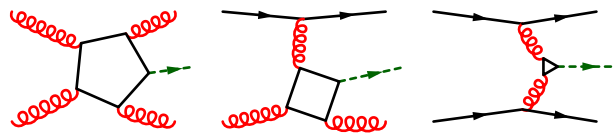

\hbox to\hsize{\hss
\includegraphics[width=.33\hsize]{fmfggh+2jet.1} 
\includegraphics[width=.33\hsize]{fmfggh+2jet.2}
\includegraphics[width=.33\hsize]{fmfggh+2jet.3}
\hss}
\caption{$H+2$~jet production via gluon fusion. Examples of Feynman
diagrams which contribute to gluon-gluon, quark-gluon and quark-quark
scattering, respectively.}
\label{fig:fig3}
\end{figure}

$H+2$~jet production via gluon fusion is known at leading order 
exactly~\cite{DelDuca:2001eu,DelDuca:2001fn}. The contributing 
subprocesses include quark-quark scattering which involves top-quark 
triangles, quark-gluon scattering which is mediated by 
triangles and boxes, and gluon-gluon scattering which requires
up to pentagon diagrams, Fig.~\ref{fig:fig3}. 
The relevant (squared) energy scales in the process $pp\rightarrow j_1j_2H$
are the parton centre-of-mass energy $s$, 
the Higgs mass $\mh^2$, the dijet invariant mass $s_{j_1j_2}$, and 
the jet-Higgs invariant masses $\sah$ and $\sbh$. At leading order they
are related through momentum conservation,
\beq
s = s_{j_1j_2} + \sah + \sbh - \mh^2\, .\label{Hjjmtmcons}
\eeq
The NLO corrections
to $H+2$~jet production via gluon fusion are not known.

In fact, in Higgs $+\ n$ jet production, up to $(n+3)$, $(n+2)$ and 
$(n+1)$--side polygon quark loops occur in gluon-gluon, quark-gluon 
and quark-quark scattering, respectively. Clearly, the complexity of
the calculation discourages one from carrying on this path. For instance,
the evaluation of the NLO corrections to 
Higgs $+$ two jet production would imply the calculation of up to
hexagon quark loops and two-loop pentagon quark loops, which are at
present unfeasible.  Fortunately, the calculations become simpler in two
instances:
\begin{itemize}
\item[-] the large $\mt$ limit, in which, as we described in Sect.~3, 
the heavy quark loop is replaced 
by an effective coupling, thus reducing the number of loops in a
given diagram by one;
\item[-] the high-energy limits, in which the number of sides in the
largest polygon quark loop is diminished at least by one.
\end{itemize}
In order to
use the large $\mt$ limit, in addition to the necessary condition
$\mh \lesssim 2 \mt$, the Higgs and jets transverse energies must be smaller
than the top-quark mass, just as it was
in the context of Higgs $+$ one jet production, Sect.~3.1. 
Then the kinematics allow us to consider two possible high energy limits: 
$(a)\ s_{j_1j_2}\gg\sah,\sbh\gg\mh^2$, \emph{i.e.} the Higgs boson is 
centrally located in rapidity between the two jets, and very far from 
either jet;
$(b)\ s_{j_1j_2},\sbh\gg\sah,\mh^2$, \emph{i.e.} the Higgs boson is 
close to jet $j_1$ in rapidity, and both of these are very far from jet $j_2$. 
In both cases the scattering amplitudes factorize into effective vertices 
connected by a gluon exchanged in the $t$ channel. Using the exact 
results~\cite{DelDuca:2001eu,DelDuca:2001fn} as a benchmark, the large 
$\mt$ limit and the high-energy limits above have been 
explored~\cite{DelDuca:2003ba}. It has been found that
the high-energy factorization is independent of the
large $\mt$ limit: the high-energy and large $\mt$ limits commute at the level
of the scattering amplitude.
In the high-energy limit, the sensitivity to the full $\mt$ 
dependence occur only locally in rapidity at the level
of the high-energy coefficient function for Higgs production.
The issue of the conditions under which it is possible to use 
the high-energy and large $\mt$ limits becomes important in
the context of its companion process: the isolation of Higgs production 
via WBF requires selecting on events with large dijet invariant mass (this cut
suppresses the gluon-gluon fusion contribution and reduces the QCD 
backgrounds). The analysis above warrants the use of the large $\mt$ limit in
gluon-gluon fusion, without regard to the value of the
Higgs--jet and/or dijet invariant masses.

\begin{figure}[h]
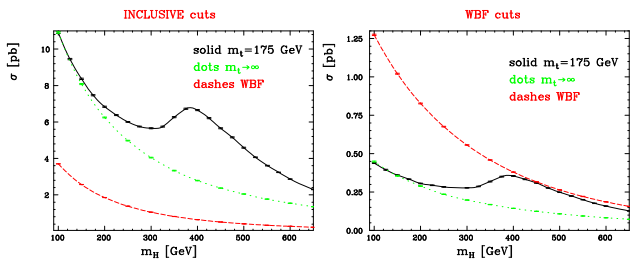

\hbox to\hsize{\hss
\includegraphics[width=.5\hsize]{ggh_no_cuts.eps} 
\includegraphics[width=.5\hsize]{ggh_cuts.eps}
\hss}
\caption{$H+2$~jet cross sections in $pp$ collisions at 
$\protect\sqrt{S}=14$~TeV as a function of the Higgs mass, from 
Ref.~\cite{DelDuca:2001eu}.
The solid and dotted lines correspond to the gluon-fusion processes 
induced by a top-quark loop
with $\mt=175$~GeV and in the large $\mt$ limit, respectively.
The dashed line corresponds to weak-boson fusion. 
The two panels correspond to two sets of jet cuts:
inclusive selection,
Eq.~(\protect\ref{eq:cuts_min}), and WBF selection, 
Eqs.~(\protect\ref{eq:cuts_min}) and (\protect\ref{eq:cut_gap}).}
\label{fig:fig4}
\end{figure}

In the left-hand side panel of Fig.~\ref{fig:fig4}, we consider $H+2$~jet
production at LHC as a function of the Higgs mass. In the exact curve for
gluon fusion, $\mt = 175$~GeV has been used (we shall use that as the default
value in the plots which follow).
On the jets, the following cuts are applied,
\beq \label{eq:cuts_min}
\ptj>20\;{\rm GeV}, \qquad |\eta_j|<5,\qquad R_{jj}>0.6,
\eeq
where $\ptj$ is the transverse momentum of a final state jet
and $R_{jj}$ describes the separation of the two partons in the 
pseudo-rapidity $\eta$ versus azimuthal angle plane.
A conspicuous feature (familiar from the inclusive gluon-fusion cross section)
of the exact $H+2$~jet gluon-fusion cross section in Fig.~\ref{fig:fig4}
is the threshold enhancement at $m_H\approx 2m_t$.
The gluon-fusion contribution dominates over the whole Higgs mass spectrum
because the cuts (\ref{eq:cuts_min}) retain events with jets in the central 
region, with relatively small dijet invariant mass.
As explained in Sect.~2, a large fraction of Higgs production via WBF 
occurs with two jets with a large rapidity interval between them.
Thus in order to isolate Higgs production via WBF from the same via
gluon-gluon fusion, we must select events with a large dijet invariant 
mass~\cite{wbfhtoww,Plehn:1999xi}. Thus, in addition to the cuts of 
Eq.~(\ref{eq:cuts_min}), on the right-hand side panel of 
Fig.~\ref{fig:fig4} we require
\beq \label{eq:cut_gap}
|\eta_{j1}-\eta_{j2}|>4.2, \quad \eta_{j1}\cdot\eta_{j2}<0, \quad
m_{jj}>600\;{\rm GeV},
\eeq
i.e. the two tagging jets must be well separated in rapidity, 
they must reside in opposite detector hemispheres and they must possess 
a large dijet invariant mass. With these selection cuts the WBF
processes dominate over gluon fusion by about 3/1 for 
Higgs masses in the 100 to 200~GeV range. 
A further suppression of gluon fusion as compared to WBF
cross sections is to be expected with a central-jet 
veto~\cite{wbfhtautau,Duff:1993ut,Barger:1994zq}, which can be
used to suppress semi-soft gluon radiation. This means that a relatively
clean separation of WBF and gluon-fusion processes will be 
possible at LHC.

As a caveat, we must note that the $H+2$~jet gluon-fusion cross section
displays a strong dependence on the renormalisation scale $\mur$: that is
because it is a leading-order calculation to $\ord(\as^4)$.
The only way to ameliorate this problem is to compute the NLO corrections.
As we discussed above, with the present-day technology an exact NLO 
calculation is not feasible. However, the NLO corrections could be
computed in the large $\mt$ limit.

\begin{figure}[h]
\hbox to\hsize{\hss
\includegraphics[width=.8\hsize]{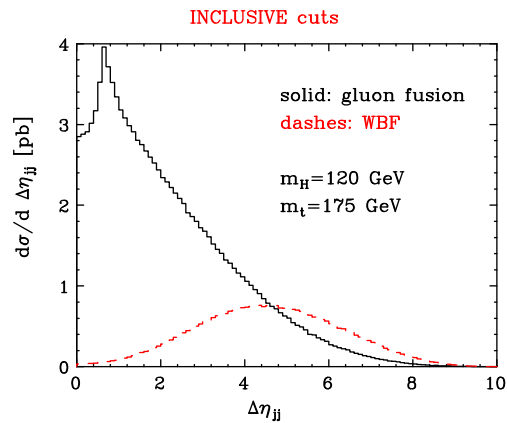} 
\hss}
\caption{Distribution in the rapidity difference $\Delta\eta_{jj}$
between the two jets, from Ref.~\cite{DelDuca:2001fn}.
The solid and dashed lines correspond to the gluon-fusion and WBF processes, 
respectively. The cuts of Eq.~(\protect\ref{eq:cuts_min}) have been used.}
\label{fig:fig5}
\end{figure}

To substantiate our claim that Higgs production via WBF tends to yield 
spontaneously two jets with a large rapidity interval between them,
in Fig.~\ref{fig:fig5} we show the distribution in the rapidity 
difference $\Delta\eta_{jj}$ between the two jets, using the cuts
of Eq.~(\protect\ref{eq:cuts_min}), for $\mh = 120$~GeV. 
While the gluon-fusion cross
section decreases monotonically (the peak at small $\Delta\eta_{jj}$ is
an artifact of the cut on $R_{jj}$), the WBF cross section produces
naturally two jets at rather large $\Delta\eta_{jj}$.

\begin{figure}[h]
\hbox to\hsize{\hss
\includegraphics[width=.8\hsize]{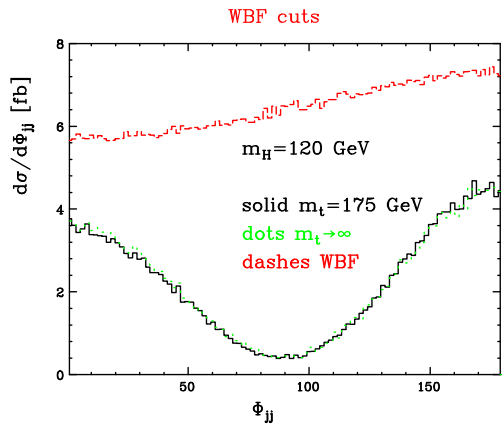} 
\hss}
\caption{Distribution in the azimuthal angle $\Delta\phi_{jj}$
between the two jets, from Ref.~\cite{DelDuca:2001fn}.
The solid and dotted lines correspond to the gluon-fusion processes 
with $\mt=175$~GeV and in the large $\mt$ limit, respectively.
The dashed line corresponds to weak-boson fusion. The cuts of
Eqs.~(\protect\ref{eq:cuts_min}) and (\protect\ref{eq:cut_gap}) 
have been used.}
\label{fig:fig6}
\end{figure}

Another discriminant between gluon fusion and WBF is the correlation
between the two jets on the azimuthal plane. It can be 
shown~\cite{DelDuca:2001fn} that while the scattering amplitudes for
WBF have a rather mild dependence on the azimuthal angle $\Delta\phi_{jj}$
between the jets, the ones for gluon fusion have an approximate zero at
$\Delta\phi_{jj} = \pi/2$. In Fig.~\ref{fig:fig6} we show the
distribution in $\Delta\phi_{jj}$\footnote{A subsequent analysis of the
distribution in $\Delta\phi_{jj}$, based on generating
the two jets and additional gluon radiation through the event
generator HERWIG~\cite{Marchesini:1983bm}, 
has found a much milder correlation in $H+2$~jet from 
gluon fusion~\cite{Odagiri:2002nd}. Unfortunately, since in that case
also the two jets were generated through the parton shower, we have no way
of comparing directly the analysis of Ref.~\cite{Odagiri:2002nd} to
Fig.~\ref{fig:fig6}.}. Note that, even with the cuts of
Eq.~(\protect\ref{eq:cut_gap}), the large $\mt$ limit
approximates very vell the exact curve, in accordance with the
discussion on the high-energy and the large-$\mt$ limits above.

\begin{figure}[h]
\hbox to\hsize{\hss
\includegraphics[width=.8\hsize]{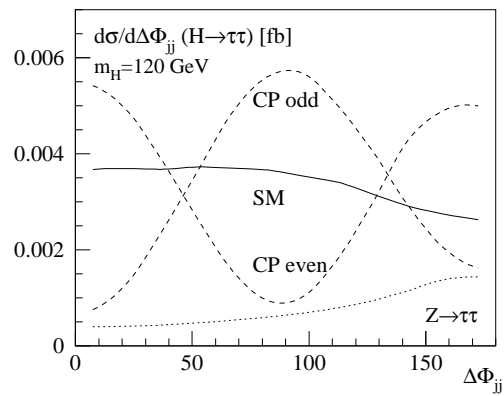} 
\hss}
\caption{Distribution in $\Delta\phi_{jj}$ for $H+2$~jet production
via WBF, with the Higgs decaying into a $\tau\tau$ pair,
from Ref.~\cite{Plehn:2001nj}.
The solid line corresponds to the Standard Model $WWH$ coupling,
the dashed lines correspond to CP-even and CP-odd anomalous couplings.
The major background $Z+2$~jets, with $Z$ decaying into $\tau\tau$, is also
represented. The cuts of
Eqs.~(\protect\ref{eq:cuts_min}) and (\protect\ref{eq:cut_gap}) 
have been used.}
\label{fig:fig7}
\end{figure}

The distribution in the azimuthal angle $\Delta\phi_{jj}$
between the two jets can be used as a tool to investigate the
tensor structure of the coupling between the Higgs and the vector 
bosons~\cite{Plehn:2001nj}. For example, let us suppose that there is
an anomalous $WWH$ coupling. This could be modelled by a
gauge-invariant effective Lagrangian, which, after expanding the Higgs
field about the v.e.v., would be given in terms of dimension 5 operators
as follows,
\begin{equation} \label{eq:anom}
{\cal L}_5 = 
  \frac{1}{\Lambda_{{\rm e}, 5}} \; H \;
    W^+_{\mu\nu} {W^-}^{\mu\nu}         
+ \frac{1}{\Lambda_{{\rm o}, 5}} \; H \; 
   \widetilde{W}^+_{\mu\nu} {W^-}^{\mu\nu}
\end{equation}
with $W^{\mu\nu}$ ($\widetilde{W}^{\mu\nu}$) the vector-boson field-strength 
(axial) tensor and $\Lambda_{{\rm e}, 5}$ ($\Lambda_{{\rm o}, 5}$)
the CP-even (odd) coupling\footnote{The Lagrangian of Eq.~(\ref{eq:anom}) 
breaks the SU(2)$\otimes$U(1) invariance, but here we are not concerned
with the consistency of the underlying fundamental theory. Rather, we want
to provide a tool to distinguish between different possible couplings,
wherever they might be come from.}. 
In Fig.~\ref{fig:fig7}, the distribution in 
$\Delta\phi_{jj}$ is shown for the SM and the anomalous
couplings, assuming that a Higgs-like scalar signal is found at LHC at the
same rate as the SM one. In this respect, $H+2$~jet from gluon fusion would
yield an undesired interference, because the effective Lagrangian which
models gluon fusion in the large $\mt$ limit is a CP-even dimension 5
operator of the same form as the one in Eq.~(\ref{eq:anom}).
Thus, there is a significant systematic uncertainty due to the poor 
determination of the normalisation of $H+2$~jet from gluon fusion,
which is only known at leading order.
The state of the affairs would be greatly improved with a calculation
of the corresponding NLO corrections.

\section*{Acknowledgments}
It is a pleasure to thank 
W.B.~Kilgore, F.~Maltoni, Z.~Nagy, C.~Oleari, C.~R.~Schmidt, Z.~Trocsanyi and 
D.~Zeppenfeld for many fruitful collaborations.

\end{document}